# Boundary Plane-Oriented Grain Boundary Model Generation

Yoyo Hinuma*, Masanori Kohyama, and Shingo Tanaka

*National Institute of Advanced Industrial Science and Technology, 1-8-31, Midorigaoka, Ikeda, Osaka 563-8577, Japan*

* y.hinuma@aist.go.jp

This study proposes algorithms for building tilt grain boundary (GB) models with a boundary plane-oriented approach that does not rely on existence of a coincidence site lattice (CSL). As conventional GB model generation uses the CSL of superimposed grains as the starting point, our totally different approach allows systematic treatment of diverse grain boundary systems that was previously not possible. Candidates of a pair of GB planes for a selected rotation axis, constituting a symmetrical or asymmetrical tilt GB, are thoroughly obtained by computational search that is applicable to any crystal structure. A GB interface for feasible computational analysis would have two-dimensional (2D) periodicity shared by the 2D lattices of the two GB planes, hence surface-slab supercells (slab-and-vacuum models) with common in-plane basis vectors of the shared 2D lattice are obtained. Finally, a procedure to obtain a GB-model supercell with alternately stacking such slabs is given. Symmetry operations of each slab may be considered such that the iterated interfaces are symmetrically equivalent, which is beneficial in *ab initio* calculations. The proposed algorithms allow streamlined generation of GB models, both symmetric and asymmetric, with or without an exact 3D-CSL relation.

## 1. Introduction

Most materials for industrial use are agglomerates of particles (grains) of a single crystal phase. Interface regions between grains are called grain boundaries (GBs), where disordered or peculiar configurations or layers are distinguished from bulk structures inside each grain.[1] GBs significantly affect various properties of materials, especially concerning diffusion[2-4], corrosion (intergranular corrosion) [5-7], fracture[8,9], and strength and deformation via affecting dislocation behaviors[10-14]. Therefore, deeply understanding GBs and their effects as well as careful design and control of GBs are beneficial not only from technological viewpoints but also as fundamental scientific topics of two-dimensional (2D) defects in solids. Atomic-level GB models are important when, for example, dealing with mechanical properties by a computational tensile test [15-17] or to identify the GB character from microscopy images[18].

Other than general GBs, there are three prototypical categories of GBs between two grains, which are tilt, twin, and twist GBs (Fig. 1). In a tilt GB (Fig. 1(a)), which is the focus of this study, the grain at one side is rotated against the other by $\theta_1+\theta_2$ as a *rotation angle*, and the rotation axis is parallel to the GB plane and shared by the two grains with the same indexes. The counterpart of one grain is its mirror image at a low index plane in a reflection twin GB [19,20](Fig. 1(b)), whereas one grain is rotated by a certain angle around a rotation axis perpendicular to the GB plane in a twist GB (Fig. 1(c)).

We hereby propose a new boundary-plane oriented approach for generation of tilt GB models for atomistic or *ab initio* simulations, which is realized by development of versatile computational techniques to build various practical symmetric or asymmetric tilt GB models with feasible two-dimensional (2D) periodic configurations. The core novelty of this work is finding a new starting point for GB model construction and demonstrating that we could attain GB models that were previously impossible, in addition to what was already possible with conventional methods, as will be discussed in detail afterwards.

Figure 2 visualizes the difference between the conventional and proposed procedures to generate tilt GB models. The conventional procedure in Fig. 2(a) constructs two superimposed grains by rotating grain B against grain A around the rotation axis. An interface plane is introduced to be parallel to the common rotation axis of the two superimposed grains. A tilt GB is completed between grains A and B by removing each half of the grains. Our new procedure (Fig. 2(b)) decides one GB plane of grain A, then a rotation axis lying on this GB plane, and finally a GB plane of grain B sharing the rotation axis. These two procedures can generate geometrically common tilt GB models because these two schemes both provide the four degrees of freedom (DOF) that determine the macroscopic geometric character of a tilt GB configuration. As a side note, there are five DOFs in a general GB[21]. In the conventional scheme using superimposed grains, two DOFs are azimuth and polar angles to determine the direction of the rotation axis in grain A, one DOF is the rotation angle of grain B around this axis, and the final DOF is associated with setting the interface plane to be parallel to the common rotation axis. For the boundary-plane oriented scheme, two DOFs fix the GB plane of grain A ($\theta_A$ and $\varphi_A$ in Fig. 2(b)), one DOF decides the rotation axis lying on this GB plane (angle $\alpha$ in Fig. 2(b)) and the final DOF sets the GB plane of grain B (angle $\beta$ in Fig. 2(b)).

The conventional superimposed-grains approach takes advantage of the three-dimensional (3D) coincident site lattice (CSL). The 3D lattices in the two grains of a GB may share a common sublattice as a CSL; a 3D-CSL relation in the superimposed grains allows relatively straightforward exploration of various symmetrical or asymmetrical GBs by setting a GB plane variously to intersect some dense or sparse 2D array of CSL points in the common 3D CSL. This is an effective procedure to find GB models with 2D periodic configurations, which may be reasonable computational models. This scheme is also applicable to GBs other than tilt GBs, depending on the relation between the rotation axis and a GB plane. There is rich literature on the CSL of a tilt GB, originally in the cubic system [22-25] and later in the hexagonal system[26]. Lists of Σ-values of tilt GBs by rotation axes, rotation angles, and GB planes are found in, for example, Refs. [27-30].

Examples of modern computational codes for GB models, which typically use to the conventional approach based on superimposed grains and the CSL, are as follows. The GBstudio software [31] is a Java applet for web browsers, where tilt and twist GB models of cubic unit cells can be built after specifying the Σ-value. More than 1000 orientations of interfaces are registered in the software, and the user chooses the desired orientation. It is also possible to specify the translation vectors by providing the rigid body translation (RBT). The GB code [32] gives possible Σ-values based on the rotation axis for cubic systems and then provides the crystal structure (either fcc, bcc, diamond, or simple cubic), the rotation axis, and a specific Σ-value to generate explicit GB models suitable for *ab initio* and molecular dynamics calculations. The aimsgb algorithm and library [29] creates tilt and twist GB models (CSL for cubic structures) when the Σ-value and rotation axis is given as input.

Unfortunately, there are many examples of GBs where an exact 3D-CSL relation in superimposed grains is impossible. The reason is not deviations from exact rotation angles, but inherent difficulty in the CSL theory for GBs in tetragonal lattices with rotation axes other than <001>, or in hexagonal lattices with rotation axes other than <0001>[33], for example, where we cannot find any 3D-CSL relation for any rotation angle other than in special cases with ideal lattice-parameter ratios. The conventional approach is not effective such situations where 3D CSL points are absent; GBs with a 2D periodic configuration cannot be searched and therefore no GB planes can be introduced into the superimposed grains. However, a GB interface can have a 2D-periodic configuration is necessary to obtain a GB model with a flat GB plane and 3D periodicity that is beneficial for *ab initio* calculations, even without an exact 3D-CSL relation, if the constituent two GB planes share some common 2D periodicity.

Our proposed algorithms can computationally find such GB plane pairs that share a common rotation axis and 2D periodicity. Symmetrical and asymmetrical tilt GBs with a 2D-periodic configuration can be obtained for any crystal system regardless of whether a 3D-CSL relation exists or not. Examples of symmetric tilt GBs where 2D-periodic configurations are constructed via the GB-plane oriented approach are discussed in Ref. [34)], namely [1$\bar{1}$00] and [1$\bar{2}$10] rotation axes in hcp metals. Superimposed grains in such a GB do not reveal any 3D array of CSL points in these systems. The indices of a GB plane (*reference plane*) of one grain are provided, then the rotation axis lying on this plane is selected, and finally we provide the counterpart GB plane (*target plane*) which shares the common rotation axis and proper in-plane 2D periodicity with the reference plane. A procedure to stack grains with symmetrically equivalent interfaces, which is useful in *ab initio* calculations, is also given.

Derivation of practical arrangement of atoms at the interface, which may involve identification of the RBT, is a substantial problem with rich complexity [35-42)] independent of choosing basis vectors of the GB model lattice, and hence is outside the scope of this study. We note that the conventional superimposed-grains approach also does not address this issue, thus the proposed procedures are not a step back from what is already accomplished. Reasonable atom position determination is a secondary problem that has meaning only after when the primary problem of GB orientation has been resolved, either by conventional methods or the proposed algorithms.

## 2. Overview of proposed algorithms

The first algorithm explained in §3.1 provides candidates of rotation axes for a given GB plane of one grain denoted as a *reference plane*. Possible rotation axes lying on the reference plane and their relative angles are specified. The minimum length of a lattice vector along the rotation axis is also provided. This is helpful in considering the 2D-lattice periodicity of the reference plane, which is essential to the construction of a tilt GB model.

The second algorithm explained in §3.2 provides possible candidates of the second GB plane of the other grain, denoted in this study as a *target plane*, for the first selected GB plane (reference plane) and rotation axis via the first algorithm. The target plane shares the same rotation axis with the reference plane. In other words, the common rotation axis is lying on both the reference and target planes. The possibility of formation of a symmetric or asymmetric tilt GB model between reference and target planes can by analyzing lattice vectors along the rotation axis in each plane. For instance, rows of such lattice vectors are periodically repeated toward the direction perpendicular to the rotation axis, and the spacing between the rows in the reference and target planes must be a rational ratio if symmetric or asymmetric tilt GB models are possible. The rotation angle decided by the two planes is also provided.

The third algorithm explained in §3.3 discusses how to construct a practical supercell of a tilt GB with selected GB planes and rotation axis via stacking of two surface slabs of the GB planes, namely the reference and target planes. There is no need of an exact 3D CSL relation between the crystals of the two slabs, while some coincidence or sharing of the 2D-lattice periodicity between the two surface slabs is needed.. In the present GB supercell, the periodicity in the direction normal to the interface is constructed by introducing a vacuum region between the stacked two slabs. We call this kind of GB supercell a slab-and-vacuum model. On the other hand, we can construct a GB supercell without any substantial vacuum region by stacking the two kinds of slabs alternately. Such a model, denoted as an alternate-stacking model, could be regarded as a special kind of slab-and-vacuum model where the vacuum layer thickness is essentially zero. The two interfaces in a GB supercell can be made symmetrically equivalent under certain conditions, which is suitable, for instance, for *ab initio* tensile tests [43-46] to obtain stress-strain curves by utilizing Nielsen-Martin cell-averaged stresses [47]. Finally, §3.4 outlines a procedure to construct alternate-stacking models with symmetrically equivalent interfaces.

The algorithms intensively use information on the spatial distribution of lattice points, while information on the atom positions is irrelevant once the lattice points are determined. Therefore, instead of the original crystal, we primarily consider a *virtual crystal* where a virtual atom is placed at each lattice point in most of the algorithms. This virtual crystal was called an *empty cell* in a previous work by Hinuma et al. [48] By using such a virtual crystal, the set of virtual atom positions can be used to describe the set of lattice vectors in the original crystal through a one-to-one relation, and the positions of lattice points can be expressed both in Cartesian coordinates and fractional coordinates in an arbitrary unit cell. Most importantly, information of lattice points can be easily tracked after changing of basis vectors, and there is no need to do a costly symmetry search after every change. Therefore, the computational time of the algorithms do not depend on the unit cell size and number of atoms once the lattice is identified. We note that atom positions in the virtual crystal are exactly the same as the original crystal in simple cubic (sc), base-centered cubic (bcc), and face-centered cubic (fcc) elementary substances. Positions of virtual atoms can be easily obtained through symmetry search software such as the spglib library [49] implemented in the phonopy code[50]. If coordinate triplets (0,0,0) and ($x$,$y$,$z$) are equivalent by virtue of translational symmetry, then there is a virtual atom at coordinate triplets ($x$,$y$,$z$) in addition to (0,0,0). The spglib library and phonopy code provide a list of all such ($x$,$y$,$z$), for an arbitrary unit cell, within the range $0 \leq x < 1$, $0 \leq y < 1$, and $0 \leq z < 1$ in fractional coordinates.

The algorithms were verified with proprietary bare-bones code, which is not available openly at this point.

## 3. Algorithms

The *basis vectors of the crystallographic conventional cell* are denoted as $(\mathbf{a},\mathbf{b},\mathbf{c})$. Vectors expressed as column vectors are usually lattice vectors connecting between lattice points in a crystal, denoted as, for example, $\mathbf{s}$. General vectors other than lattice vectors are denoted with a breve as, for instance, $\breve{\mathbf{n}}$.

### 3.1 Finding rotation axes in a reference plane

This section finds rotation axis candidates when a GB plane is given. In other words, lattice vectors are searched in a lattice plane (reference plane) with given indices. Denoting the indices of the reference plane and rotation axis as $\left(h_{\text{ref}} k_{\text{ref}} l_{\text{ref}}\right)$ and $\left[h_{\text{rot}} k_{\text{rot}} l_{\text{rot}}\right]$, respectively, the following relation holds:

$$\left(h_{\text{ref}}, k_{\text{ref}}, l_{\text{ref}}\right) \cdot \left(h_{\text{rot}}, k_{\text{rot}}, l_{\text{rot}}\right) = 0 . \tag{1}$$

Numbers in indices are not separated by a comma, while commas are used to separate components in a three-dimensional vector. This is a trivial problem for crystals without centering; one only needs to scan $h_{rot}$, $k_{rot}$, and $l_{rot}$ over a predetermined range. The values $h_{rot}$, $k_{rot}$, and $l_{rot}$ are chosen to be coprime, which means that the positive greatest common divisor among non-zero values of $h_{rot}$, $k_{rot}$, and $l_{rot}$ is 1. For example, $(h_{rot}, k_{rot}, l_{rot}) = (-2, -1, 0)$ is considered coprime, but $(h_{rot}, k_{rot}, l_{rot}) = (-2, 0, 0)$ is not. The combination $(h_{rot}, k_{rot}, l_{rot}) = (0, 0, 0)$ does not describe a meaningful direction, therefore is not considered.

The next problem is to find a lattice vector with minimum length along the rotation axis (rotation vector, denoted with the symbol $s$ from "shared"). The rotation vector is a key quantity that becomes a basis vector for both the reference plane and the other GB plane (target plane). Its length is critical when trying to minimize the size of the GB model used in calculations; a model with a shorter rotation vector results in a smaller GB unit cell under periodic boundary conditions and therefore smaller computational cost. The shortest length of a lattice along the rotation axis when there is no centering is simply

$$|h_{rot}\mathbf{a} + k_{rot}\mathbf{b} + l_{rot}\mathbf{c}|. \tag{2}$$

However, searching the shortest length of a lattice along the rotation axis is not trivial in crystals with centering. A procedure is given in Supplementary Material 1[51]) together with an example.

Figure 3 shows an example of a search for rotation vectors in NaC$_2$. This material was chosen because its Bravais lattice is simple cubic (space group $P4_332$). The pink plane in Fig. 3 is the (011) plane. The number, species, and positions of atoms on the lattice does not matter at all in searching for rotation vectors; only the lattice points (intersection of black lines) are relevant. There are eight rotation vector candidates with maximum index 1, which are shown in white arrows in Fig. 3. As expected from eq. (2), the length of the rotation vectors are 1, $\sqrt{2}$, and $\sqrt{3}$ for rotation vectors <100>, <01$\bar{1}$>, and <11$\bar{1}$>, respectively, when the basis vector lengths are scaled to $|\mathbf{a}| = |\mathbf{b}| = |\mathbf{c}| = 1$. There are two, two, and four such vectors, respectively, in Fig. 3.

*3.2 Identification of target plane candidates after the reference plane and rotation axis are given*

The proposed algorithm obtains candidates of the second GB plane (*target plane*, indices $\left(h_{targ} k_{targ} l_{targ}\right)$), after the first GB plane (*reference plane*, indices $\left(h_{ref} k_{ref} l_{ref}\right)$), and the *rotation axis* (indices $\left[h_{rot} k_{rot} l_{rot}\right]$) are specified for a given crystal. The angle between the reference and target planes is the rotation angle.

The algorithm also provides information on the 2D-lattice periodicities of the reference and target planes, which is especially important when checking whether parts of the lattice points of the two planes share a 2D lattice. The term "CSL" is generally reserved for the shared 3D lattice between two grains forming a GB, thus we use the term "shared 2D lattice" for our purpose. As actual examples, the rutile (221) $\left[1\bar{1}0\right]$ GB of rutile $TiO_2$ [18] and $\left[1\bar{1}00\right]$ and $\left[1\bar{2}10\right]$ symmetric tilt boundaries of hcp Mg[34] each have a 2D shared lattice but an exact 3D CSL does not exist.

The algorithm in this section is independent from that in §3.1. Information on the actual atom positions is irrelevant in this section, and therefore we work on the virtual crystal (§2) with virtual atoms on lattice points.

A simple way to find target plane candidates is to, in a procedure similar to §3.1, scan $h_{targ}$, $k_{targ}$, and $l_{targ}$ over a certain range to find a combination of coprime $h_{targ}$, $k_{targ}$, and $l_{targ}$ satisfying

$$\left(h_{targ}, k_{targ}, l_{targ}\right) \cdot \left(h_{rot}, k_{rot}, l_{rot}\right) = 0. \tag{3}$$

The vector

$$\breve{\boldsymbol{n}} = h\mathbf{b}\times\mathbf{c} + k\mathbf{c}\times\mathbf{a} + l\mathbf{a}\times\mathbf{b} \tag{4}$$

is perpendicular to the (*hkl*) plane. Therefore, using vectors normal to the reference and target planes,

$$\breve{\boldsymbol{n}}_{\text{ref}} = h_{\text{ref}}\mathbf{b}\times\mathbf{c} + k_{\text{ref}}\mathbf{c}\times\mathbf{a} + l_{\text{ref}}\mathbf{a}\times\mathbf{b} \tag{5}$$

and

$$\breve{\boldsymbol{n}}_{\text{targ}} = h_{\text{targ}}\mathbf{b}\times\mathbf{c} + k_{\text{targ}}\mathbf{c}\times\mathbf{a} + l_{\text{targ}}\mathbf{a}\times\mathbf{b}, \tag{6}$$

it is possible to obtain the angle between the reference and target planes $\theta$ in the range $0° \leq \theta \leq 180°$ through

$$\theta = \cos^{-1}\frac{\breve{\boldsymbol{n}}_{\text{ref}} \cdot \breve{\boldsymbol{n}}_{\text{targ}}}{|\breve{\boldsymbol{n}}_{\text{ref}}||\breve{\boldsymbol{n}}_{\text{targ}}|}. \tag{7}$$

This "rotation angle $\theta$" in the present study can be used to easily obtain the "inclination angle $\Phi$" in Ref. [30, 52], which is defined as the angle around the rotation axis between a pair of GB planes with a certain rotation relation.

Unfortunately, the above procedure does not provide information of the periodicity of the 2D lattices of the reference and target planes. Analysis of the periodicity is facilitated by explicitly finding lattice vectors that act as basis vectors of a specific form.

An *in-plane reference vector* $\boldsymbol{r}_{\text{IP}}$, a *rotation vector* $\boldsymbol{s}$, and an *out-of-plane reference vector*, $\boldsymbol{r}_{\text{OP}}$ are chosen as a set of primitive basis vectors defining the 3D lattice of the crystal with handedness

$$(\boldsymbol{r}_{\text{IP}} \times \boldsymbol{s}) \cdot \boldsymbol{r}_{\text{OP}} > 0. \tag{8}$$

Vectors $\boldsymbol{r}_{\text{IP}}$ and $\boldsymbol{s}$ define the 2D lattice of the reference plane in this setup, and use of these vectors allows us to easily identify all lattice points on one side of the reference plane as a linear combination of integers times basis vectors. Similarly, an *in-plane target vector* $\boldsymbol{t}_{\text{IP}}$ and the rotation vector $\boldsymbol{s}$ are chosen to be primitive basis vectors of the 2D lattice of the target plane. The rotation vector $\boldsymbol{s}$ is therefore shared by both the reference and target planes. Vectors $\boldsymbol{s}$, $\boldsymbol{r}_{\text{IP}}$, $\boldsymbol{r}_{\text{OP}}$ are obtained using the procedure in §S2.1 in Supplementary Material 2[51]. One out-of-plane target vector that is not in the

target plane is

$$t'_{OP} = h_{targ}\mathbf{a} + k_{targ}\mathbf{b} + l_{targ}\mathbf{c}. \tag{9}$$

This vector does not necessarily act as a basis vector that forms primitive cell with $t_{IP}$ and $s$.

A target plane can be uniquely defined using an in-plane target vector

$$t_{IP} = p\mathbf{r}_{IP} + q\mathbf{r}_{OP}. \tag{10}$$

and $s$, where $p$ and $q$ are coprime integers and $q \geq 0$, and $p=1$ only if $q=0$ (target plane is the same as the reference plane). Expressing this in-plane target vector in terms of the original basis vectors as

$$t_{IP} = h'_{targ\_IP}\mathbf{a} + k'_{targ\_IP}\mathbf{b} + l'_{targ\_IP}\mathbf{c}, \tag{11}$$

the indices of the target plane are defined through

$$(h_{targ}, k_{targ}, l_{targ}) = m(h'_{targ\_IP}, k'_{targ\_IP}, l'_{targ\_IP}) \times (h_{rot}, k_{rot}, l_{rot}). \tag{12}$$

where $m$ is a positive number that makes $h_{targ}$, $k_{targ}$, and $l_{targ}$ coprime. Taking the number of atoms in the conventional cell as $M$, which is one of 1, 2, 3, and 4 depending on the centering, $h'_{targ\_IP}$, $h_{rot}$, etc. are integer multiples of $1/M$. Therefore, $h_{targ}$, $k_{targ}$, and $l_{targ}$ are integer multiples of $1/M^2$, hence $m$ is $M^2$ divided by a positive integer.

There is a search in three-dimensional parameter space in eq. (3), namely $h_{targ}$, $k_{targ}$, and $l_{targ}$. In contrast, there is a scan over two variables, $p$ and $q$, in eq. (10). Variables $h_{targ}$, $k_{targ}$, and $l_{targ}$ cannot be determined independently because of the constraint in eq. (3) that removes one DOF among $h_{targ}$, $k_{targ}$, and $l_{targ}$. Indices $(h_{targ}k_{targ}l_{targ})$ represent the target plane in the conventional cell with basis vectors $(\mathbf{a}, \mathbf{b}, \mathbf{c})$, while the indices are $(q0\bar{p})$ in the primitive cell with basis vectors $(\mathbf{r}_{IP}, s, \mathbf{r}_{OP})$.

We now explore the periodicity of the 2D lattice in the reference and target planes.

The periodicity in the direction perpendicular to the rotation axis on each plane is important when screening for possibility of a shared 2D lattice between the two planes. Figure 4 illustrates the complexity of the problem. Examples of 2D lattices of a reference plane, constituted by rotation and in-plane reference vectors, are shown. In a rectangular (or square) lattice oriented as shown in Fig. 4(a), the distance between the rotation vectors in the normal direction ($d_{ref}$ in Fig. 4) is equal to the in-plane reference vector length, $|\mathbf{r}_{IP}|$. However, this is not the case in, for example, an oblique lattice (Fig. 4(b)), where

$$d_{ref} = \left| \mathbf{r}_{IP} - \frac{\mathbf{r}_{IP} \cdot \mathbf{s}}{\mathbf{s} \cdot \mathbf{s}} \mathbf{s} \right|. \tag{13}$$

The second term is zero when the rotation and in-plane reference vectors are perpendicular to each other. This distance $d_{ref}$ is important in examining a possible 2D shared lattice. Figures 4(c) and 4(d) are two planes of the same crystal sharing the same rotation vector $\mathbf{s}$ (black arrows), and the yellow arrows have the same length and angle with respect to $\mathbf{s}$. Therefore, the yellow and black arrows become basis vectors of identical 2D lattices in Figs. 4(c) and 4(d). As a result, a GB with a 2D shared lattice can be formed if Figs. 4(c) and 4(d) are regarded as the reference and target planes. The ratio of $d_{ref}$ in Figs. 4(c) and 4(d) is rational (3:1); a rational ratio of $d_{ref}$ between two planes sharing $\mathbf{s}$, strictly a rational ratio between $d_{ref}$ and its counterpart in the target plane that is mentioned later, $d_{targ}$, is a necessary condition for the existence of a 2D shared lattice. In contrast, the ratio of $|\mathbf{r}_{IP}|$ and its counterpart in the target plane is not directly concerned with the 2D shared lattice.

There are an infinite number of target planes. One way of reducing candidates to be considered is to apply the condition that a similar distance between the rotation vectors in the target plane,

$$d_{targ} = \left| \mathbf{t}_{IP} - \frac{\mathbf{t}_{IP} \cdot \mathbf{s}}{\mathbf{s} \cdot \mathbf{s}} \mathbf{s} \right|. \tag{14}$$

is below a certain cutoff (preferably a given constant times the minimum value, $d_{targ\_min}$). Target planes are then sorted by the rotation angle $\theta$ between reference and target planes, or to be precise, the angle between vectors perpendicular to the reference and target planes, $\breve{\boldsymbol{n}}_{\text{ref}}$ and $\breve{\boldsymbol{n}}_{\text{targ}}$, respectively (eqs. (5) and (6), respectively). Equation (7) gives $\theta$ in the range $0° \leq \theta \leq 180°$, but defining the angle $\theta$ against the reference plane in the range $0° \leq \theta < 360°$ is much more convenient, as discussed later. The necessary procedure is given in Supplementary Material 3[51]. In essence, the rotation angle is considered as the azimuth angle of $\breve{\boldsymbol{n}}_{\text{targ}}$ in cylindrical coordinates when vector $\breve{\boldsymbol{n}}_{\text{ref}}$ is taken as the reference, or $\varphi = 0$, and the cylindrical direction is taken along $s$, which is perpendicular to both $\breve{\boldsymbol{n}}_{\text{ref}}$ and $\breve{\boldsymbol{n}}_{\text{targ}}$. The above procedure investigated half of the lattice vectors in the *pq*-space in eq. (10); adding values of $(-h_{\text{targ}}, -k_{\text{targ}}, -l_{\text{targ}})$ for each $(h_{\text{targ}}, k_{\text{targ}}, l_{\text{targ}})$ results in a list of all target vectors; in the former, $d_{\text{targ}}$ is the same as the latter and $\theta$ is different by $180°$. A table of $h_{\text{targ}}$, $k_{\text{targ}}$, $l_{\text{targ}}$, $d_{\text{targ}}$, optionally $d_{\text{targ}}$ normalized by $d_{\text{targ\_min}}$, and $\theta$ (an example is Table I) would be convenient for further analysis.

The above algorithm is demonstrated on fcc Cu in Fig. 5. The reference plane is (111) (pink plane in Fig. 5) and the rotation axis is $[0\bar{1}1]$ (intersection of pink and blue planes in Fig. 5). Table I shows a list of indices of target planes and rotation angles with respect to the (111) reference plane of fcc Cu with rotation axis is $[0\bar{1}1]$.

*3.3. Constructing a tilt GB with shared 2D-lattice periodicity*

This section discusses how to construct a GB model with a shared 2D-lattice based on the reference and target plane indices.

The *indices of the reference and target planes* are given, as input, as $(h_{\text{ref}} k_{\text{ref}} l_{\text{ref}})$ and $(h_{\text{targ}} k_{\text{targ}} l_{\text{targ}})$. How to obtain the $\Sigma$-value and indices of GB plane pairs (both symmetric and asymmetric) from the rotation axis and angle when there is a CSL is given in Supplementary Material 5[51] to provide a stronger link to existing literature. There exists a procedure to derive the (*hkl*)-primitive cell with basis vectors $(\mathbf{a}_\mathrm{P}, \mathbf{b}_\mathrm{P}, \mathbf{c}_\mathrm{IP})$, which is a primitive cell of bulk where the first two basis vectors constitute the (*hkl*) plane as *in-plane basis vectors* [48]. A supercell of the form

$$(\mathbf{a}', \mathbf{b}', \mathbf{c}') = (\mathbf{a}_\mathrm{P}, \mathbf{b}_\mathrm{P}, \mathbf{c}_\mathrm{IP}) \begin{pmatrix} * & * & * \\ * & * & * \\ 0 & 0 & * \end{pmatrix}, \quad (15)$$

where asterisks * represent integers and all the asterisks do not need to be the same number, has the characteristic that $\mathbf{a}'$ and $\mathbf{b}'$ are in-plane basis vectors. Then a slab with the (*hkl*) orientation can be obtained by removing atoms with *z*-coordinates outside a given range. We call this kind of a surface-slab supercell a slab-and-vacuum model. A GB model is constructed by using the slab-and-vacuum models of the reference and target planes, where the in-plane basis vectors in the $(h_\mathrm{ref} k_\mathrm{ref} l_\mathrm{ref})$ and $(h_\mathrm{targ} k_\mathrm{targ} l_\mathrm{targ})$ planes should constitute the 2D shared lattice.

The algorithm in this section consists of three steps. The first step is to obtain basis vectors of the 2D lattice shared by the two planes. Here, there is a need to check whether reference and target planes can share a 2D lattice. Next, *reference and target base cells* are obtained that describe slab-and-vacuum models using readily available in-plane basis vectors in the reference and target planes. Finally, supercells of the base cells with the same periodicity of the 2D shared lattice are constructed as *reference and target matching cells*, respectively.

*3.3.1 Obtaining basis vectors of the shared 2D lattice*

We start from the crystallographic conventional cell with basis vectors $(\mathbf{a}, \mathbf{b}, \mathbf{c})$. Primitive vectors of the 2D reference plane are denoted as the *reference in-plane vector* $\boldsymbol{r}_\mathrm{IP}$ and *rotation vector* $\boldsymbol{s}$, and those of the 2D target plane as the *target in-plane vector* $\boldsymbol{t}_\mathrm{IP}$ and $\boldsymbol{s}$. Note that $\boldsymbol{s}$ appears as a primitive basis vector in both reference and target planes. These vectors are obtained using the procedure in §S6.1[51] (the derivation is slightly different from §S2.1).

The lattice vectors $r_{IP}$ and $t_{IP}$ are decomposed into two vectors each:

$$\breve{r}_{//s} = \left( \frac{r_{IP} \cdot s}{s \cdot s} \right) s \tag{16}$$

$$\breve{r}_{\perp s} = r_{IP} - \breve{r}_{//s}, \tag{17}$$

and

$$\breve{t}_{//s} = \left( \frac{t_{IP} \cdot s}{s \cdot s} \right) s \tag{18}$$

$$\breve{t}_{\perp s} = t_{IP} - \breve{t}_{//s}. \tag{19}$$

Here, $\breve{r}_{//s}$ and $\breve{t}_{//s}$ are projections of $r_{IP}$ and $t_{IP}$ onto $s$, respectively, and $\breve{r}_{\perp s}$ and $\breve{t}_{\perp s}$ are both perpendicular to $s$. The perpendicular and parallel components of a lattice vector are not necessarily lattice vectors, for example in an oblique 2D lattice. The perpendicular component is always non-zero, while the parallel component may be occasionally zero, which could be the case in a non-centered rectangular or square lattice.

The 2D lattices of the reference and target planes, constituting the GB interface, share some 2D-lattice periodicity if:

-Two coprime natural numbers $m$ and $n$ exist such that

$$m|\breve{r}_{\perp s}| = n|\breve{t}_{\perp s}|. \tag{20}$$

-Two coprime integers $p$ and $q$, where $q>0$, exist such that

$$m\breve{r}_{//s} - n\breve{t}_{//s} = (p/q)s. \tag{21}$$

All of $m$, $n$, and $q$ are 1 in a symmetric GB, but other values are possible in asymmetric GBs. A procedure to convert a decimal number into a reasonable fraction (rational number) is given in Supplementary Material 7[51].

We find a set of basis vectors in the reference plane, $\tilde{r}_{IP}$ and $s$, and those in the target plane, $\tilde{t}_{IP}$ and $s$, that are primitive basis vectors of the shared 2D lattice when rotated appropriately. A lattice vector in the reference plane,

$$\tilde{r}_{IP} = mq(\breve{r}_{\perp s} + \breve{r}_{//s}) + us, \qquad (22)$$

can be rotated around the rotation axis by an appropriate angle to obtain a lattice vector in the target plane,

$$\tilde{t}_{IP} = nq(\breve{t}_{\perp s} + \breve{t}_{//s}) - ps + us. \qquad (23)$$

Here, $u$ is an arbitrary integer.

$$\tilde{r}_{IP} = mqr_{IP} - \lfloor 0.5 + mq(\breve{r}_{//s} \cdot s)/(s \cdot s) \rfloor s, \qquad (24)$$

can be rotated to obtain a lattice vector in the target plane,

$$\tilde{t}_{IP} = nqt_{IP} - ps - \lfloor 0.5 + mq(\breve{r}_{//s} \cdot s)/(s \cdot s) \rfloor s. \qquad (25)$$

The floor function term ($\lfloor \cdot \rfloor$) acts to minimize the norms of $\tilde{r}_{IP}$ and $\tilde{t}_{IP}$.

We now have information on the in-plane basis vectors of the matching cells, and now we proceed with making actual base and matching cells.

*3.3.2 Base cell determination*

Deriving the $(h_{ref}k_{ref}l_{ref})$-primitive cell with in-plane basis vectors $\mathbf{a}_{P\_ref}$ and $\mathbf{b}_{P\_ref}$ is a simple problem. A slab-and-vacuum model with a (*hkl*) surface can be obtained by making a $1 \times 1 \times N$ supercell of the (*hkl*)-primitive cell, or (*hkl*) *N*-supercell, [48] and removing atoms outside the range $z_- < z < z_+$ ($z_-$ and $z_+$ are chosen, as appropriate, between 0 and 1). In the slab-and-vacuum model, the same slabs are repeated with vacuum regions inserted as in a usual surface-slab supercell, and the out-of-plane vector is adjusted to control the vacuum-region thickness and stacking manner of slabs. In this case, the out-of-plane vector does not have to be a lattice vector of the original lattice.

Unfortunately, basis vectors $\mathbf{a}_{P\_ref}$ and $\mathbf{b}_{P\_ref}$ are not necessarily the same as $\tilde{r}_{IP}$ and $s$. For example, for a primitive cubic cell with basis vectors $(\mathbf{a},\mathbf{b},\mathbf{c})$, the basis vectors of the (001)-primitive cell are simply $(\mathbf{a},\mathbf{b},\mathbf{c})$. However, when we make a GB with the [110] rotation axis, $s$ becomes $\mathbf{a}+\mathbf{b}$, which is not one of the basis vectors of the (001)-primitive cell. We therefore need a transformation matrix to bridge between a unit cell with in-plane basis vectors $\mathbf{a}_{P\_ref}$ and $\mathbf{b}_{P\_ref}$ and another unit cell with in-plane basis vectors $\tilde{r}_{IP}$ and $s$.

The reference base cell is hereby defined as a slab-and-vacuum model with basis vectors

$$\left(\mathbf{a}_{P\_ref}, \mathbf{b}_{P\_ref}, \breve{r}_{OP\_match}\right), \tag{26}$$

where the in-plane basis vectors $\mathbf{a}_{P\_ref}$ and $\mathbf{b}_{P\_ref}$ are the same as the $\left(h_{ref} k_{ref} l_{ref}\right)$-primitive cell. This cell acts as a base for construction of the reference matching cell. The third basis vector, $\breve{r}_{OP\_match}$, is an arbitrary out-of-plane vector that is recommended to be orthogonal to the in-plane vectors, including all of $\mathbf{a}_{P\_ref}$, $\mathbf{b}_{P\_ref}$, $\tilde{r}_{IP}$, and $s$. This orthogonality requirement is not an intrinsic one, but is assumed for the following procedure. Atoms outside some ranges of $z$ (normal direction of each plane) are removed to make slab-and-vacuum models. As mentioned above, the size of $\breve{r}_{OP\_match}$ is decided to properly set a vacuum-region thickness.

Figure 6 illustrates the procedure to make a reference base cell using CsCl (space group $Pm\bar{3}m$). The position of virtual atoms overlaps with those of Cs. The (111)-primitive cell with (111) as the *ab*-plane and the (111) 4-supercell are shown in Figs. 6(a) and 6(b), respectively. Then atoms outside the range $0.2 < z < 0.7$ are removed in Fig. 6(c). In Fig. 6(d), the out-of-plane basis vector $4\mathbf{a}$ in Fig. 6(c) is changed to $\breve{r}_{OP\_match} = 1.5(\mathbf{a}+\mathbf{b}+\mathbf{c})$, which is not a lattice vector of the original lattice, with fixed atomic positions in the slab denoted by the orange arrow.

*3.3.3 Matching cell determination*

The basis vectors of the reference matching cell with the shared 2D lattice, which is a supercell of the reference base cell, are denoted as

$$\left(\tilde{\boldsymbol{r}}_{\text{IP}}, \boldsymbol{s}, \breve{\boldsymbol{r}}_{\text{OP\_match}}\right). \tag{27}$$

The goal is to make two matching cells with the same shared 2D lattice, one with basis vectors $\tilde{\boldsymbol{r}}_{\text{IP}}$, $\boldsymbol{s}$, and $\breve{\boldsymbol{r}}_{\text{OP\_match}}$, and the other with $\tilde{\boldsymbol{t}}_{\text{IP}}, \boldsymbol{s}$, and $\breve{\boldsymbol{t}}_{\text{OP\_match}}$ defined similarly. This allows the transfer of fractional internal coordinates from one matching cell to another to obtain a GB model.

The reference matching cell is constructed as a supercell of the reference base cell using the transformation matrix $\mathbf{M}_{\text{ref}}$ defined by

$$\left(\tilde{\boldsymbol{r}}_{\text{IP}}, \boldsymbol{s}, \breve{\boldsymbol{r}}_{\text{OP\_match}}\right) = \left(\mathbf{a}_{\text{P\_ref}}, \mathbf{b}_{\text{P\_ref}}, \breve{\boldsymbol{r}}_{\text{OP\_match}}\right)\mathbf{M}_{\text{ref}}, \tag{28}$$

or after rearrangement of the equation,

$$\mathbf{M}_{\text{ref}} = \left[\left(\mathbf{a}_{\text{P\_ref}}, \mathbf{b}_{\text{P\_ref}}, \breve{\boldsymbol{r}}_{\text{OP\_match}}\right)^{-1}\left(\tilde{\boldsymbol{r}}_{\text{IP}}, \boldsymbol{s}, \breve{\boldsymbol{r}}_{\text{OP\_match}}\right)\right]. \tag{29}$$

If $\tilde{\boldsymbol{r}}_{\text{IP}} \cdot \breve{\boldsymbol{r}}_{\text{OP\_match}} = 0$ and $\boldsymbol{s} \cdot \breve{\boldsymbol{r}}_{\text{OP\_match}} = 0$, the transformation matrix is guaranteed to take the form

$$\mathbf{M}_{\text{ref}} = \begin{pmatrix} * & * & 0 \\ * & * & 0 \\ 0 & 0 & 1 \end{pmatrix} \tag{30}$$

where the asterisks * represent integers and different asterisks may take different values. Such a transformation matrix is necessary to transform coordinate triplets of atom positions from one cell to another.

The target matching cell with basis vectors $\tilde{\boldsymbol{t}}_{\text{IP}}, \boldsymbol{s}$, and $\breve{\boldsymbol{t}}_{\text{OP\_targ}}$ is obtained similarly. First, a target base cell with basis vectors

$$\left(\mathbf{a}_{\text{P\_targ}}, \mathbf{b}_{\text{P\_targ}}, \breve{\boldsymbol{t}}_{\text{OP\_match}}\right). \tag{31}$$

is obtained, where the in-plane basis vectors $\mathbf{a}_{\text{P\_targ}}$ and $\mathbf{b}_{\text{P\_targ}}$ are the same as the $\left(h_{\text{targ}} k_{\text{targ}} l_{\text{targ}}\right)$-primitive cell and the out-of-plane vector $\breve{\boldsymbol{t}}_{\text{OP\_match}}$ is decided in a similar way to $\breve{\boldsymbol{r}}_{\text{OP\_match}}$. The transformation matrix $\mathbf{M}_{\text{targ}}$ between the target base and matching cells satisfies

$$\left(\tilde{\boldsymbol{t}}_{\mathrm{IP}}, s, \breve{\boldsymbol{t}}_{\mathrm{OP\_match}}\right) = \left(\mathbf{a}_{\mathrm{P\_targ}}, \mathbf{b}_{\mathrm{P\_targ}}, \breve{\boldsymbol{t}}_{\mathrm{OP\_match}}\right) \mathbf{M}_{\mathrm{targ}}. \tag{32}$$

Once the two matching cells with the same shared 2D-lattice periodicity are constructed, a slab-and-vacuum model of a GB supercell is formed by stacking the two slabs.

*3.3.4 Example of matching cell derivation*

A demonstration of matching cell derivation is given in Supplementary Material 8.[51)] Figure 7 shows the $\left(\bar{1}11\right)$ and $\left(\bar{5}11\right)$ planes of fcc Cu (pink and blue, respectively) as well as important vectors appearing in the algorithm. Figures 8(a) and 8(b) shows supercells where the in-plane basis vectors are those of the $\left(\bar{1}11\right)$- and $\left(\bar{5}11\right)$-primitive cells as defined in Hinuma et al.[48)]. Two choices of the $\left(\bar{1}11\right)$-primitive cell exists, one with $\gamma=60°$ and another with $120°$, and the latter is adopted in this study. There are eight and 24 layers, respectively, in each cell. Although the out-of-plane basis vector is taken to be as orthogonal to the in-plane basis vectors as possible, it is not exactly orthogonal in both cases. The number of atom layers is reduced to three and nine in Figs. 8(c) and 8(d), respectively, and the out-of-plane basis vector is taken to be exactly orthogonal to the in-plane basis vectors. Supercells of Figs. 8(c) and 8(d) built using these transformation matrices are shown in Figs. 8(e) and 8(f), respectively. The lattice parameters of these supercells are the same, as intended, which is useful for GB model generation.

*3.4. Building a GB-model supercell with two equivalent GBs*

*3.4.1 Conditions for stacked slabs to generate symmetrically equivalent interfaces*

This section considers an alternate-stacking model of a GB supercell where two kinds of slabs, 1 and 2, are alternately stacked without any vacuum layer (examples are shown in Fig. 9(a)). There are two interfaces in a unit cell; interface A between the lower side of slab 2 and the upper side of slab 1, and interface B between the lower side of slab 1' and the upper side of slab 2. Models where these two interfaces are symmetrically equivalent are often desirable. The simple way to attain such models, even for asymmetric-GB cases, is the "symmetrized slab scheme" where both sides of each slab are made symmetrically equivalent by a common symmetry operation shared by the two slabs. The out-of-plane vector is obtained by connecting symmetry elements in a certain manner. There are an infinite number of choices, but a shorter out-of-plane vector is often more favorable because it is more orthogonal to the interface plane.

The present symmetrized slab scheme is not possible in the case of polar surfaces of compounds because both sides of the slab have different stoichiometry. A supercell with two equivalent interfaces with arbitrary surfaces, including polar surfaces, can be constructed if slab 2 is an image of slab 1 due to a two-fold screw with a screw-rotation axis perpendicular to the interfaces or a glide reflection with a reflection plane perpendicular to the interfaces (an example is found in Ref. [53]). This approach is not pursued further in this study.

There are five types of symmetry operations on a slab of a bulk crystal where the image, after operation, of the upper side of the slab is the lower side and vice versa: inversion, two-fold or two-fold screw rotations with respect to an axis parallel to the slab-surface plane, and mirror or glide-reflection operations with respect to a plane parallel to the slab-surface plane. Whether two sides of an isolated slab of a bulk crystal are symmetrically equivalent or not is not necessarily evident from eyeballing, and identifying symmetry operations in a slab by some robust scheme is necessary. For each slab configuration, the symmetry elements, in matrix-vector notation[54], can be extracted by symmetry analysis code, for example, the spglib library[49] implemented in the phonopy code[50].

The case of inversion symmetry is discussed first. Figure 9(a) shows the positions of inversion centers in a GB supercell of alternate-stacking slabs with symmetrically equivalent interfaces. Red circles represent inversion centers in slabs 1 and 1' while blue circles show inversion centers in slabs 2 and 2'. The out-of-plane basis vector (blue arrow) defines the relative positions of slabs 1' and 2 against slabs 1 and 2', respectively, by translation. We select an inversion center each in slabs 1 and 2, and the out-of-plane vector is simply double this connecting vector. Here, the RBT represents the freedom of relative translation between the slabs, and the RBT of interface A is defined as the translation of slab 2 against slab 1. There is no restriction on this RBT of interface A, but the RBT of interface B, defined as the relative translation of slab 1' against slab 2, is determined uniquely from the RBT of interface A to keep the inversion symmetry.

Figure 9(b) shows an example of mirror symmetry parallel to the interface in each slab. Red planes indicate mirror planes of slabs 1 and 1', while blue planes show those of slabs 2 and 2'. The distance between adjacent mirror planes must be the same in the supercell to make all the interfaces symmetrically equivalent. The out-of-plane basis vector originates at an arbitrary point on the mirror plane center in slab 1 and terminates at its image in slab 1' generated by the mirror plane in slab 2 (blue arrow in Fig. 9(b)). Therefore, the $c$-axis must be normal to the $ab$-plane (interaxial angles $\alpha=\beta=90°$). The RBT of each interface can have components parallel to the interface in addition to the normal component. Importantly, the RBT component parallel to the mirror plane can be taken independent of the out-of-plane vector.

Two-fold rotation symmetry is discussed last, where the rotation axes are arbitrarily taken parallel to the $b$-axis as in Figs. 9(c) and 9(d), where views from two different directions are shown. The symmetry requires that the $a$- and $c$-axis must be normal to the $b$-axis (interaxial angles $\alpha=\gamma=90°$), while there is no restriction on interaxial angle $\beta$. Red arrows indicate two-fold rotation axes of slabs 1 and 1', while blue arrows show those of slabs 2 and 2'. We pick two points, one in slab 1 and the other in slab 2, as a crossing point between a two-fold rotation axis and an $ac$-plane normal to the $b$-axis. The coordinate of the $ac$-plane along the $b$-axis may be selected arbitrarily. Now we consider the connecting vector between these two points in slabs 1 and 2, and the out-of-plane vector is double this connecting vector (blue arrow in Figs. 9(c) and 9(d)). The RBTs of interfaces A and B have the following properties. The symmetry does not restrict the RBT at interface A, defined as the shift of slab 2 against slab 1. In contrast, as for the RBT of interface B, defined as the shift of slab 1' against slab 2, the RBT components on the $ac$-plane have to be the same as those of interface A, while the RBT component parallel to the $b$ axis has to be opposite to that of interface A. Thus, the total sum of the RBTs of the two interfaces has components on the *ac*-plane only, which is contained in the out-of-plane vector.

*3.4.2 Examples of building alternate-stacking models*

This section discusses generation of alternate-stacking models of MgO for GB models with a 2D shared lattice interface. In the present example shown in Fig. 10, the reference and target planes are (100) and (001), respectively, and the rotation axis is [010]. This is a Σ1 tilt GB with rotation angle 90°. The (100) and (001) slabs are labeled as slabs 1 and 2 in this section. Although the models obtained here are hypothetical or unrealistic as a GB, especially as the RBT is not optimized, the objective is to demonstrate how to build an alternate-stacking model.

MgO has a lattice parameter of 4.2 Å. A three-layer slab is shown in Fig. 10(a), with internal coordinates shown in Table II. The relative atom positions within the two slabs are identical, therefore only one set is shown in Fig. 10(a). The atoms in the center layer are inversion centers (one of them is emphasized with a circle), two-fold rotation axes parallel to the *b*-axis penetrate atoms in the center layer (one is shown as an arrow), and the center layer is a mirror plane (rectangle). There are other symmetry elements, but are set aside for now. Slab 2 is translated by (1.3 Å, 0.8 Å, 7.0 Å) from slab 1 (purple arrows) in Figs. 10(b)-10(d). When the two interfaces are made symmetrically equivalent by inversion, the out-of-plane basis vector *c* connecting slab 1 with its image, slab 1', becomes (2.6 Å, 1.6 Å, 14.0 Å) (Fig. 10(b)). This is double the vector between inversion centers of slabs 1 and 2. In contrast, when the symmetry element is mirror (Fig. 10(c)), the *c* basis vector is (0 Å, 0 Å, 14.0 Å), which is normal to the *ab*-plane. When two-fold rotation parallel to the *b*-axis is the symmetry element, the *c* basis vector is (2.6 Å, 0 Å, 14.0 Å) (Fig. 10(d)).

Figure 11 shows further examples of symmetrized slab scheme GB models generated using the algorithms in this study. Detailed information used to generate the models, such as the unit cell and relevant vectors, are given in Supplementary Material 9 [51]. RBTs, which are defined here as the shift of atoms in the target matching cell (gray color) with regard to those of the reference matching cell (orange color), were arbitrarily determined and are not optimized with respect to GB formation energy. Figure 11(a) is an asymmetric hcp Mg $(1\bar{1}0)(5\bar{8}0)[001]$ GB model. The RBT was chosen such that the GB model satisfies the symmetrized slab scheme while keeping the out-of-plane vector normal to the GB plane. The out-of-plane vector in Fig. 11(a) is clearly not a basis vector of the matching cell, thus this model cannot be obtained from the CSL. Using the algorithms in this study therefore provides significant flexibility on the choice of basis vectors of the GB model while ensuring that the model satisfies the symmetrized slab scheme. Figures 11(b) and 11(c) are symmetric Mg $(\bar{3}21)(\bar{2}3\bar{1})[111]$ and base-centered tetragonal (bct) In $(\bar{4}11)(\bar{1}4\bar{1})[113]$ GB models, respectively. The is no CSL for these combinations of crystal family and rotation axis. There is also no RBT for Fig. 11(b). The out-of-plane vector of the GB model is not normal to the GB plane although the out-of-plane vector of the matching cells are normal (Fig. S4(b)) because the former is determined from the connecting vector of inversion centers in the matching cells. The RBT for Fig. 11(c) contains non-zero values in three dimensions, and a symmetrized slab scheme GB model was successfully obtained in such a complicated situation.

## 4. Summary

Four algorithms related to atomic-level tilt GB model generation were proposed in this study. These algorithms are focused on building models from information on indices of GB planes and existence of a 2D-lattice periodicity at the GB interface, while conventional methods require a much stricter requirement, namely a 3D-coincidence relation between rotated crystals. There are numerous GB systems where exact 3D-CSL relation is inherently impossible, thus the algorithms greatly extend the variety of GB model construction. Rotation axis candidates can be identified from the orientation of a GB plane (reference plane) based on §3.1. Candidates for the other GB plane (target plane) could be obtained from the initially selected GB plane and rotation axis indices according to §3.2. Section 3.3 discussed how to construct a practical supercell of a tilt GB with selected GB planes and rotation axis, via stacking of two slabs with reference and target GB planes. The proposed algorithm examines the 2D-lattice periodicity of each reference and target GB planes, and based on the shared 2D-lattice periodicity, a slab-and-vacuum model of a GB supercell is constructed. Finally, §3.4 outlined a procedure to construct alternate-stacking models with symmetrically equivalent interfaces, suitable for *ab initio* calculations.

The significance of our results lies in expansion of the horizons of GB model generation. The developed tools allow making of GB models that were not previously possible systematically. The conventional approach, although better than none, confined us to the tiny set of crystal structure and rotation axis systems where the 3D CSL relation must hold. Combinations where the crystal structure and rotation axis can never satisfy the 3D CSL relation is the norm rather than the exception. Finding matching GBs under the restriction of a 2D periodicity at the interface, regardless of symmetric or asymmetric, opens up numerous possibilities and avenues of grain boundary research.

**Acknowledgment**

This study was funded by a grant (No. JPMJCR17J3) from CREST of the Japan Science and Technology Agency (JST) and JSPS KAKENHI 21H05101. The VESTA code [55] was used to draw Figs. 3, 5-11, S1 and S4.
*E-mail: y.hinuma@aist.go.jp

Table I. List of indices of target planes (*hkl*) and rotation angles (θ) with respect to the (111) reference plane of fcc Cu. The rotation axis is $[0\bar{1}1]$, and entries are limited to $d_{\text{targ}}/d_{\text{targ\_min}} < 4$.

| h | k | l | $d_{\text{targ}}$ (Å) | $d_{\text{targ}}/d_{\text{targ\_min}}$ | θ (°) |
|---|---|---|---|---|---|
| 1 | 1 | 1 | 2.20 | 1 | 0 |
| 5 | 3 | 3 | 8.35 | 3.79 | 14.42 |
| 2 | 1 | 1 | 6.24 | 2.83 | 19.47 |
| 3 | 1 | 1 | 4.22 | 1.91 | 29.50 |
| 5 | 1 | 1 | 6.61 | 3 | 38.94 |
| 1 | 0 | 0 | 2.55 | 1.15 | 54.74 |
| 5 | -1 | -1 | 6.61 | 3 | 70.53 |
| 3 | -1 | -1 | 4.22 | 1.91 | 79.98 |
| 2 | -1 | -1 | 6.24 | 2.83 | 90 |
| 5 | -3 | -3 | 8.35 | 3.79 | 95.05 |
| 1 | -1 | -1 | 2.20 | 1 | 109.47 |
| 1 | -2 | -2 | 7.64 | 3.46 | 125.26 |
| 1 | -3 | -3 | 5.55 | 2.52 | 131.47 |
| 0 | -1 | -1 | 3.60 | 1.63 | 144.74 |
| -1 | -3 | -3 | 5.55 | 2.52 | 158.00 |
| -1 | -2 | -2 | 7.64 | 3.46 | 164.21 |
| -1 | -1 | -1 | 2.20 | 1 | 180 |
| -5 | -3 | -3 | 8.35 | 3.79 | 194.42 |
| -2 | -1 | -1 | 6.24 | 2.83 | 199.47 |
| -3 | -1 | -1 | 4.22 | 1.91 | 209.50 |
| -5 | -1 | -1 | 6.61 | 3 | 218.94 |
| -1 | 0 | 0 | 2.55 | 1.15 | 234.74 |
| -5 | 1 | 1 | 6.61 | 3 | 250.53 |
| -3 | 1 | 1 | 4.22 | 1.91 | 259.98 |
| -2 | 1 | 1 | 6.24 | 2.83 | 270 |
| -5 | 3 | 3 | 8.35 | 3.79 | 275.05 |
| -1 | 1 | 1 | 2.20 | 1 | 289.47 |
| -1 | 2 | 2 | 7.64 | 3.46 | 305.26 |
| -1 | 3 | 3 | 5.55 | 2.52 | 311.47 |
| 0 | 1 | 1 | 3.60 | 1.63 | 324.74 |
| 1 | 3 | 3 | 5.55 | 2.52 | 338.00 |
| 1 | 2 | 2 | 7.64 | 3.46 | 344.21 |

Table II. Internal coordinates ($x,y,z$), in Cartesian coordinates, of a MgO (100) and (001) three-layer slab. The lattice parameter of MgO is 4.2 Å. The basis vectors of the slab are $\mathbf{a}' = (4.2, 0, 0)^T$ and $\mathbf{b}' = (0, 4.2, 0)^T$. The unit is Å.

| Atom | $x$ | $y$ | $z$ |
|---|---|---|---|
| Mg | 0.1 | 0.1 | 0.1 |
| Mg | 2.2 | 2.2 | 0.1 |
| Mg | 0.1 | 2.2 | 2.2 |
| Mg | 2.2 | 0.1 | 2.2 |
| Mg | 0.1 | 0.1 | 4.3 |
| Mg | 2.2 | 2.2 | 4.3 |
| O | 0.1 | 2.2 | 0.1 |
| O | 2.2 | 0.1 | 0.1 |
| O | 0.1 | 0.1 | 2.2 |
| O | 2.2 | 2.2 | 2.2 |
| O | 0.1 | 2.2 | 4.3 |
| O | 2.2 | 0.1 | 4.3 |

Fig. 1. Schematic images of tilt, reflection twin, and screw grain boundaries. The cross in a circle indicates the position of the rotation axis, which is perpendicular to the page.

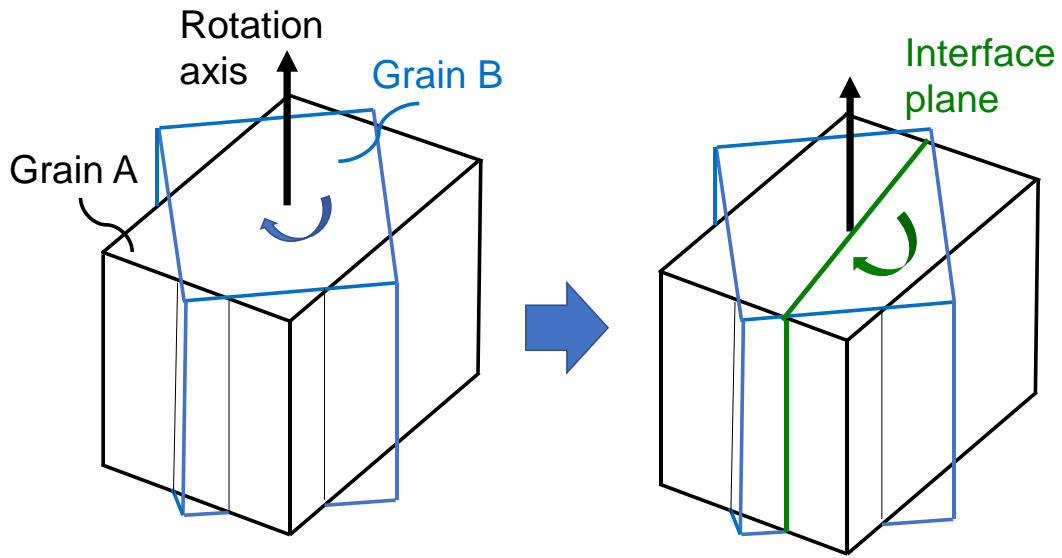

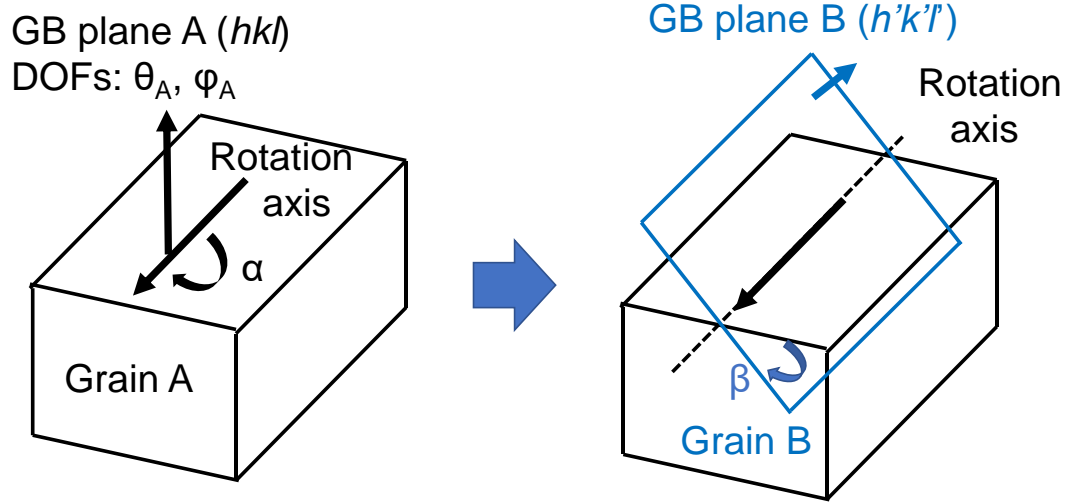

Fig. 2. Schematic on how to obtain a GB model based on a (a) conventional and (b) GB plane-oriented procedure.

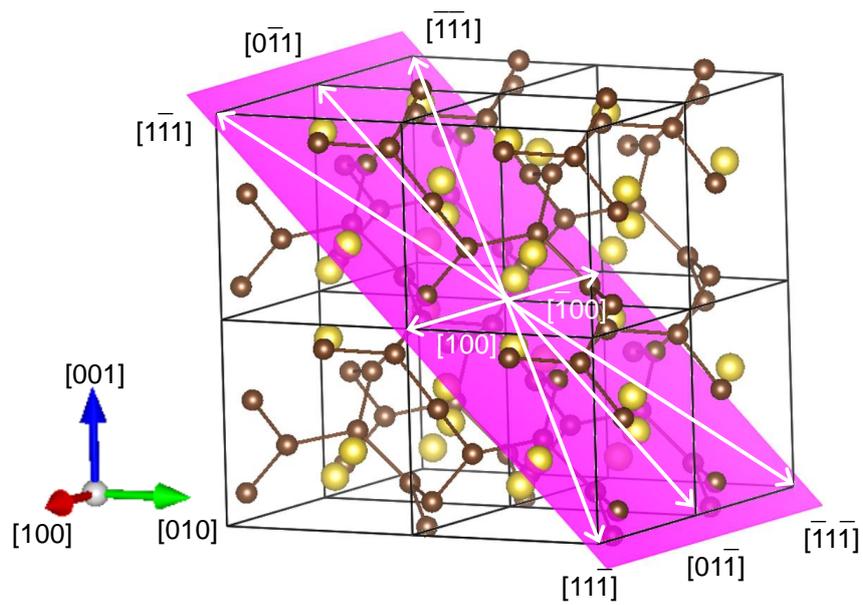

Fig. 3. Rotation vector candidates (white arrows) on the (011) reference plane (pink plane) of simple cubic NaC$_2$. Yellow and brown balls indicate Na and C, respectively.

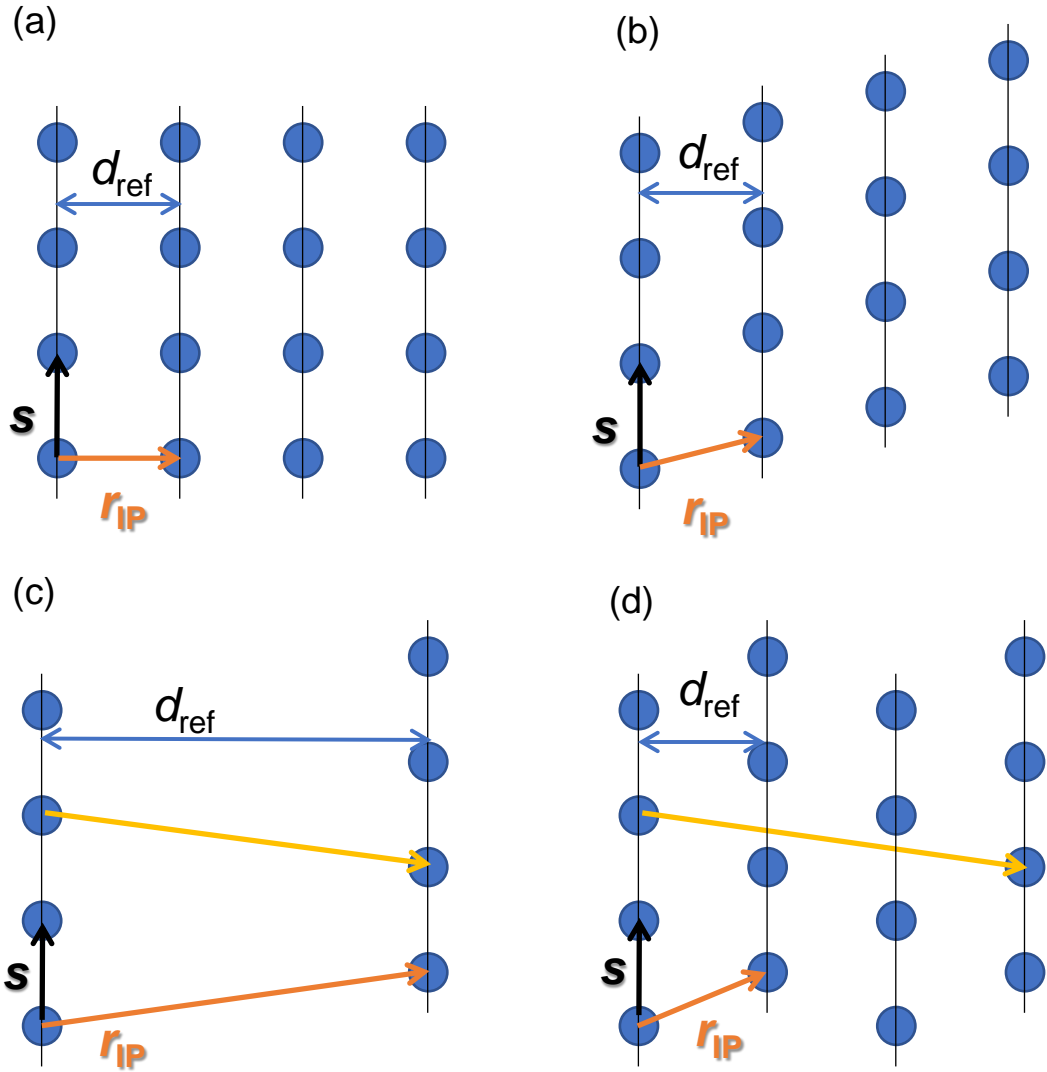

Fig. 4. Relation between the rotation vector $s$, in-plane rotation vector $r_{IP}$, and distance between rows of atoms (blue circles) parallel to the rotation axis (black lines), $d_{ref}$, in a reference plane. (a) Rectangular lattice, (b) oblique lattice and (c,d) centered lattices with different periodicity.

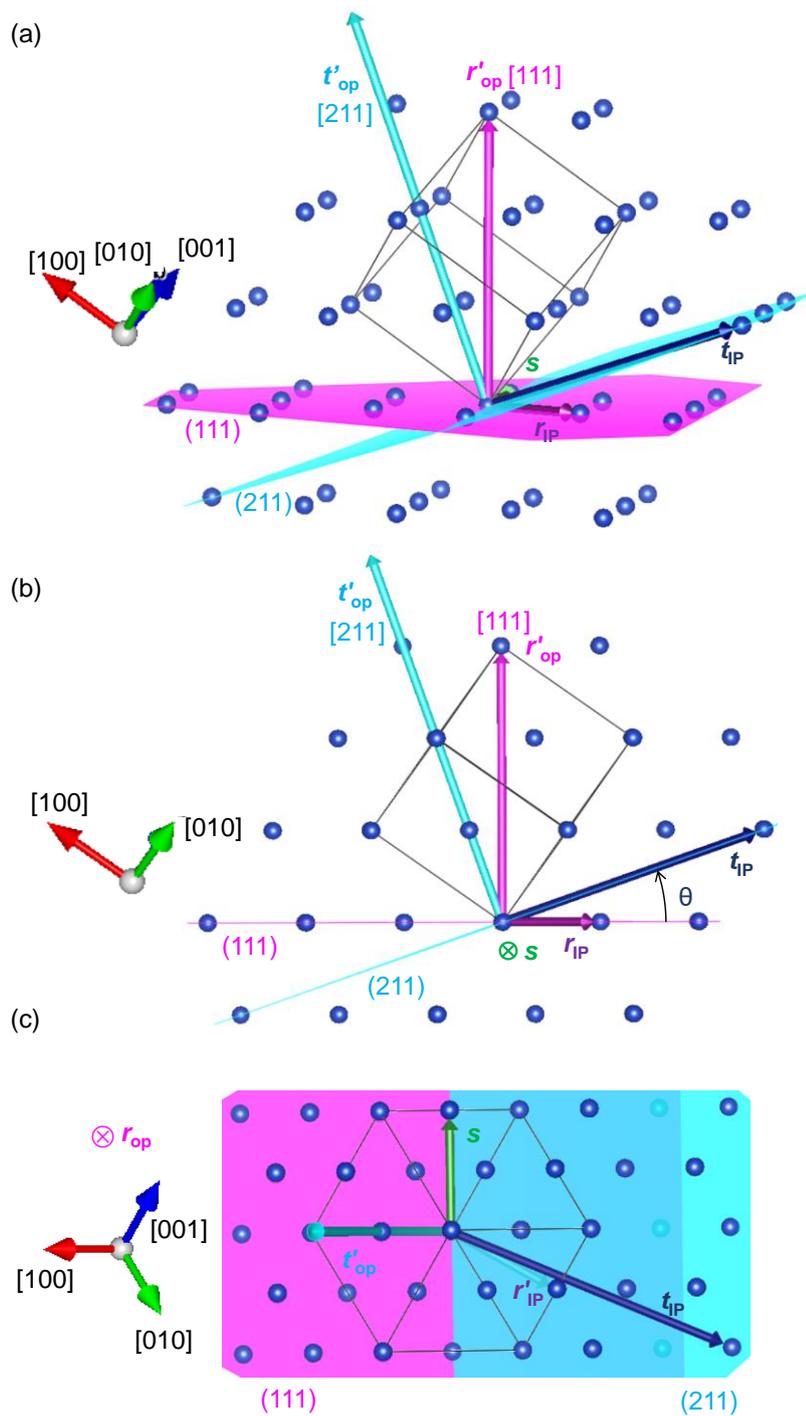

Fig. 5. Derivation of a target GB plane, which is (211), from the (111) reference plane in fcc Cu. The derivation is given in §S4. Important vectors are shown, and the same information is drawn from three directions. The definition of the angle θ between the reference and target plane is shown in (b); note that the rotation vector *s* (direction $[0\bar{1}1]$) is pointing into the page in (b).

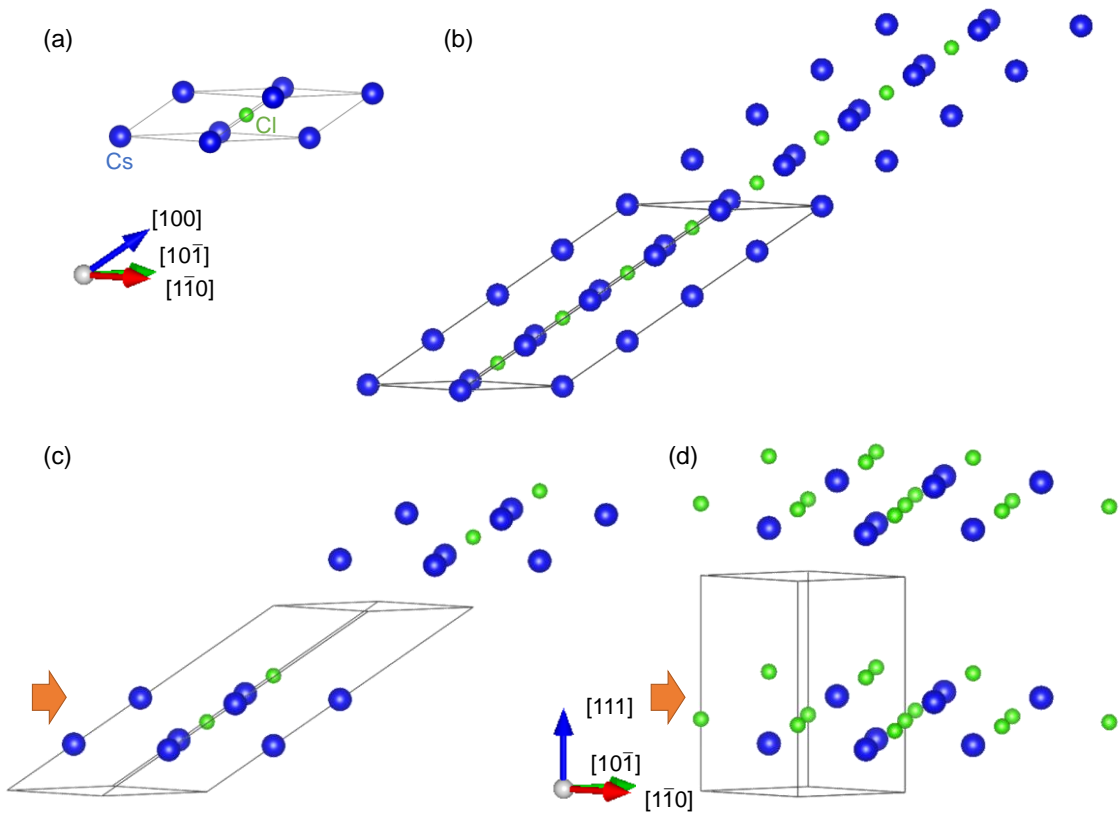

Fig. 6. Making a (111) slab of primitive cubic CsCl. Large blue and small green circles indicate Cs and Cl atoms, respectively. (a) (111)-primitive cell [48]. (b) (111) 4-supercell, which is an 1×1×4 supercell of the (111)-primitive cell. (c) Slab-and-vacuum model obtained by removing atoms from the (111) 4-supercell. (d) The out-of-plane basis vector is retaken to be normal to the (111) plane in the vacuum layer; the resulting model is a slab-and-vacuum model although the out-of-plane basis vector is not a lattice vector of the original lattice.

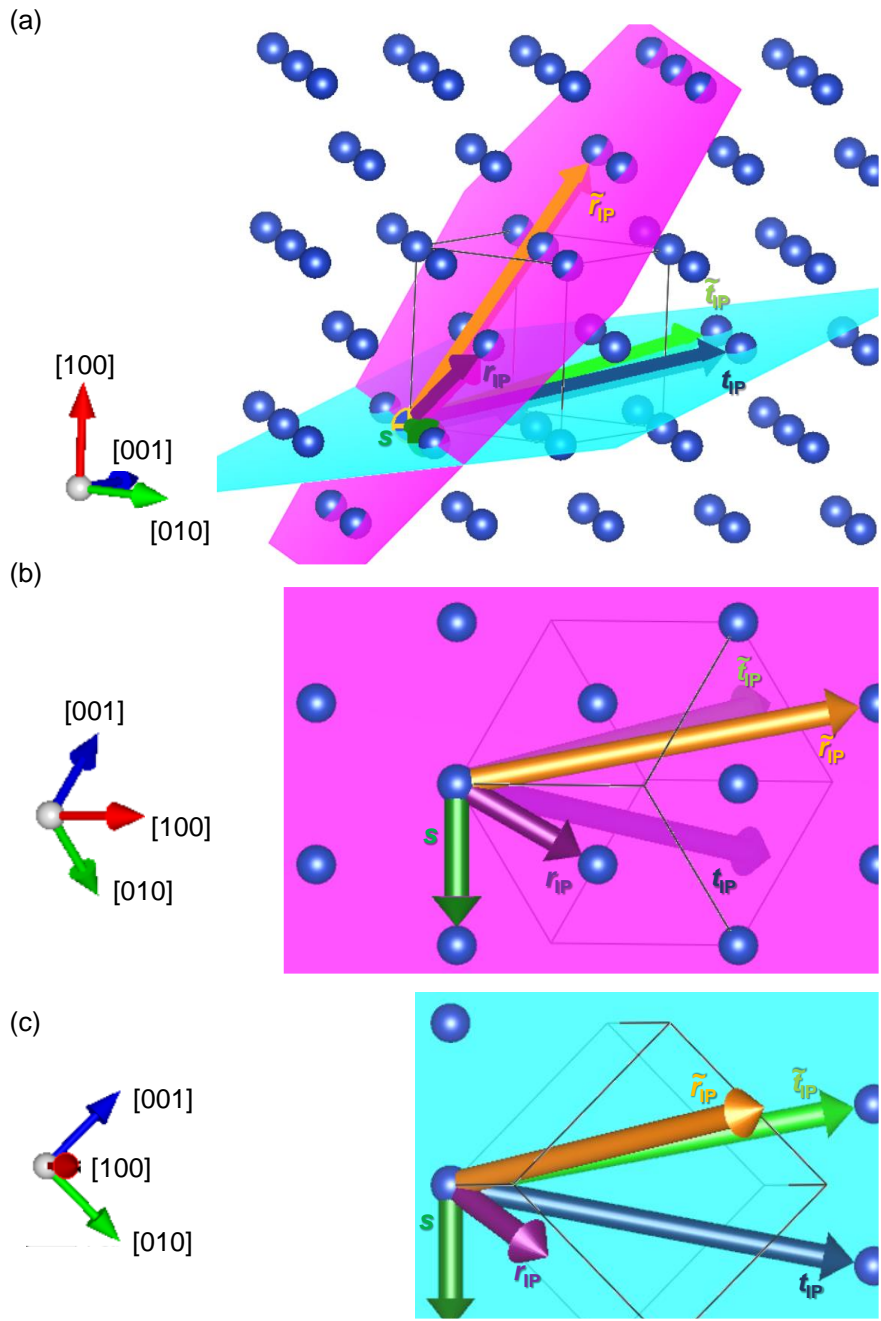

Fig. 7. Important vectors appearing during the derivation in the algorithm to check the possibility of asymmetrical tilt GB formation between the $(\bar{1}11)$ reference plane (pink) and the $(\bar{5}11)$ target plane (blue) in fcc Cu. The derivation is given in §S8. The same information is drawn from three directions.

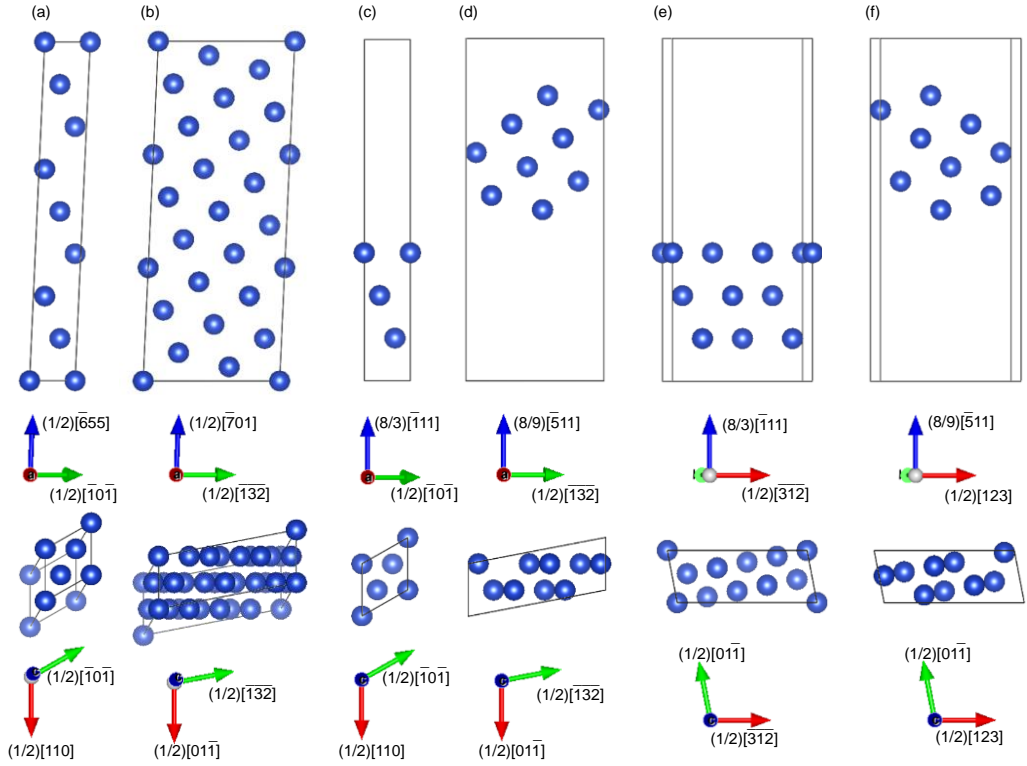

Fig. 8. (a) 1×1×8 supercell of the $(\bar{1}11)$-primitive cell with the out-of-plane basis vector taken to make it as orthogonal as possible to the in-plane basis vectors. (b) Similar 1×1×24 supercell of the $(\bar{5}11)$-primitive cell. (c) Supercell in (a) with layers of atoms reduced to three and the out-of-plane basis vector is chosen to be precisely orthogonal to the in-plane basis vectors. (d) Supercell in (b) changed similarly; the number of layers is nine. (e) Supercell of (c) obtained using the transformation matrix in eq. (S80). (f) Supercell of (d) obtained using the transformation matrix in eq. (S83).

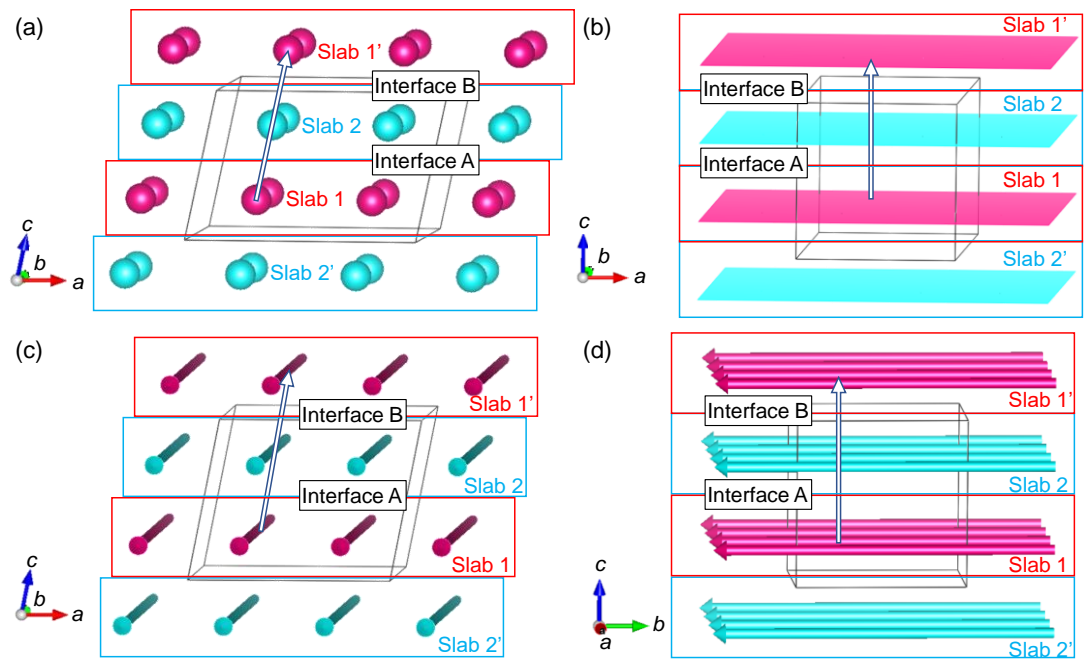

Fig. 9. Symmetry elements in an alternate-stacking model. Pink and blue motifs indicate symmetry elements in a different slab. The symmetry elements are (a) inversion, (b) mirror parallel to the *ab*-plane, and (c,d) two-fold rotation parallel to the *b* axis, shown from two directions. The *a*-and *b*-axes are parallel to an interface, while the *c*-axis is not. All eight inversion centers, two mirror planes, and four rotation axes are shown in a primitive cell (for details on the positions of symmetry elements, see the entry for $P\bar{1}$, *Pm*, and *P*2 space groups, respectively, in the International Tables of Crystallography A[56]).

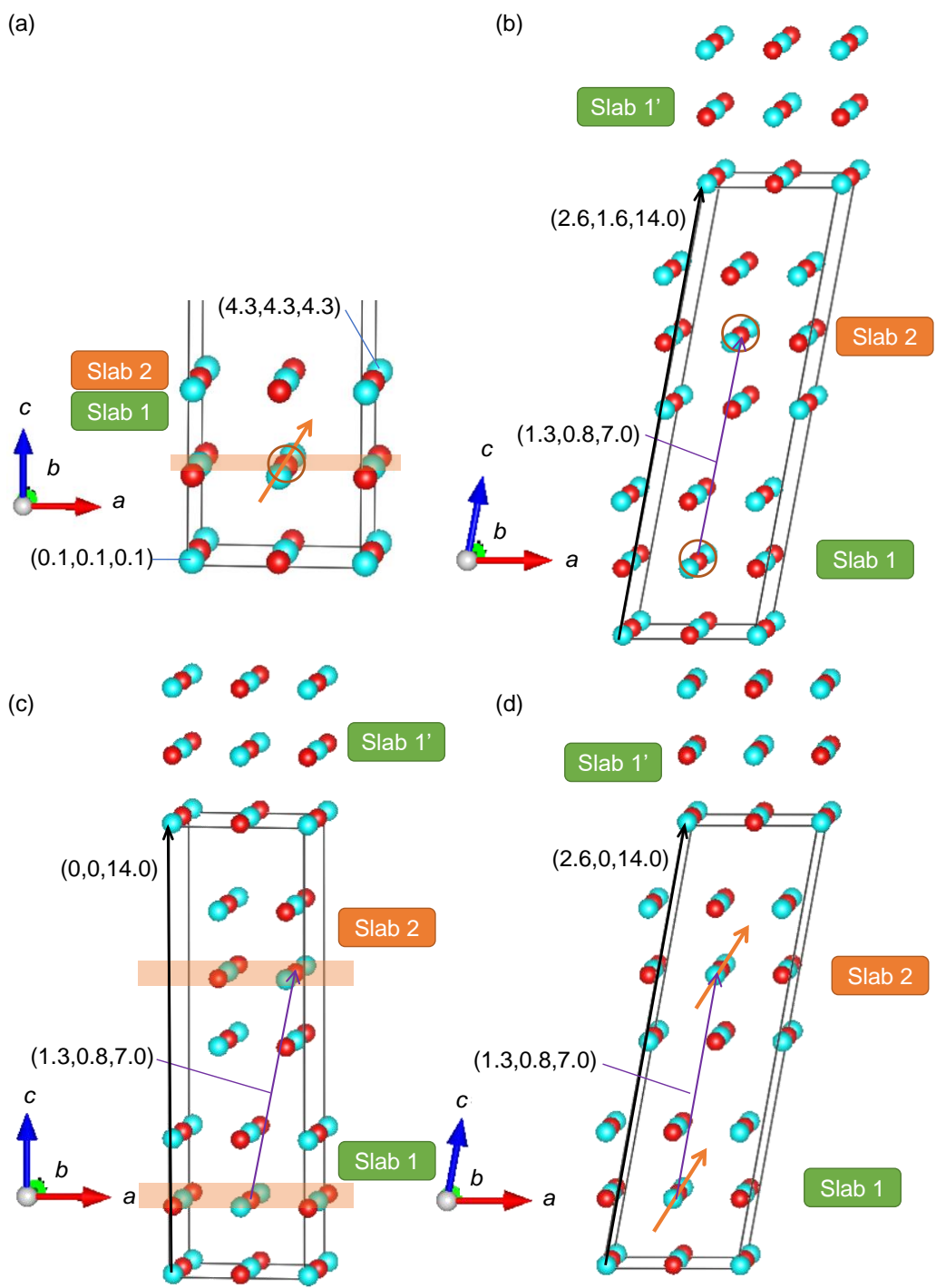

Fig. 10. Making alternating-stacking models with symmetrically equivalent interfaces. (a) MgO slab with (100) or (001) surface. (b) Model with inversion symmetry (orange circles). (c) Model with mirror symmetry (orange planes). (c) Model with two-fold rotation symmetry (orange arrows).

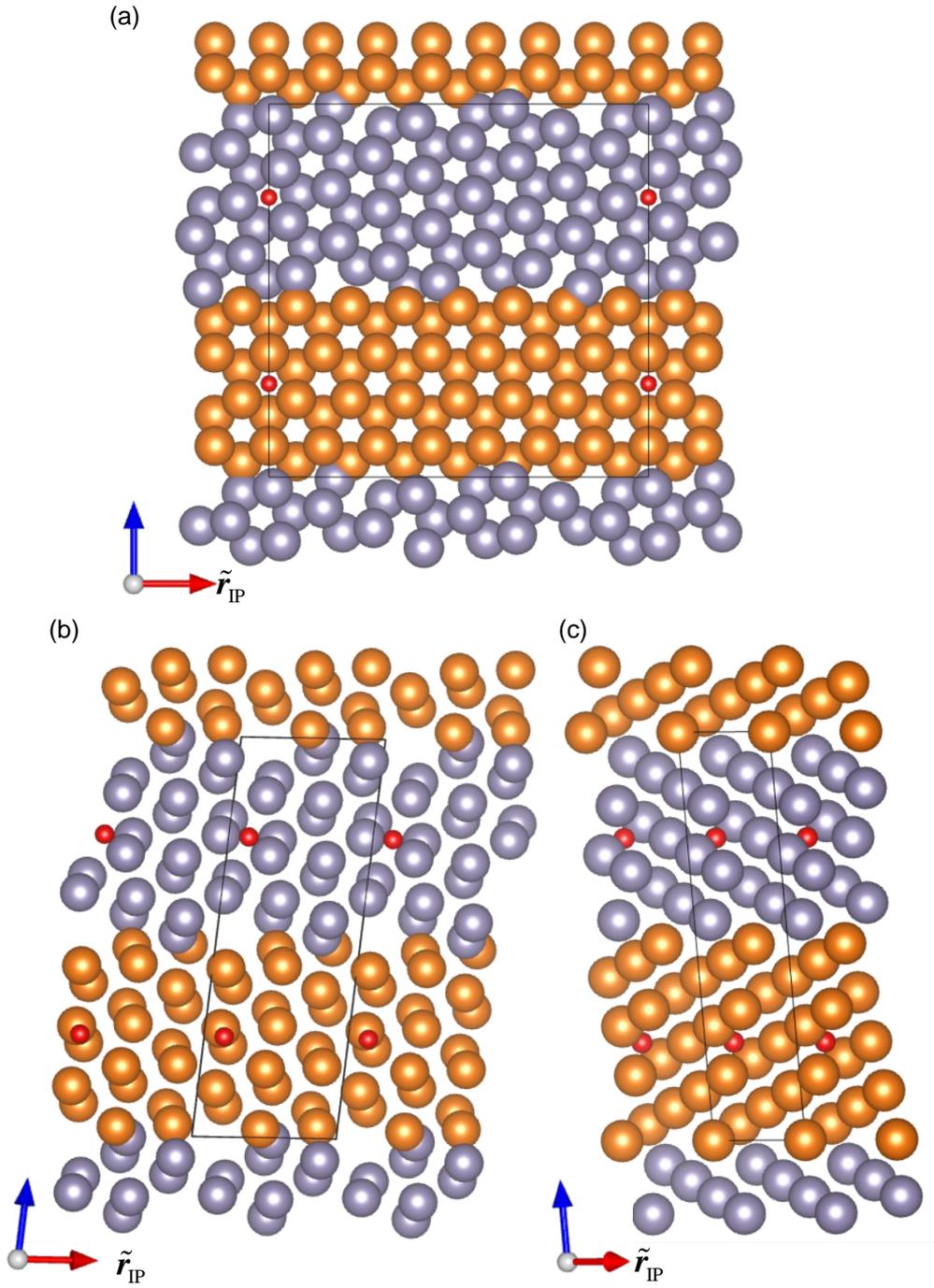

Fig. 11. GB models of (a) hcp $\mathrm{Mg}(1\bar{1}0)(5\bar{8}0)[001]$, (b) $\mathrm{Mg}(\bar{3}21)(\bar{2}3\bar{1})[001]$, and (c) bct $\mathrm{In}(\bar{4}11)(\bar{1}4\bar{1})[113]$. Information used to obtain these models is given in Supplementary Material 9[51]. Red points indicate inversion centers that ensure the symmetrized slab scheme. The rotation vector $s$ points into the page.